\documentclass[letterpaper]{article} 
\usepackage{multirow}
\usepackage{aaai2026}  
\usepackage{times}  
\usepackage{helvet}  
\usepackage{courier}  
\usepackage[hyphens]{url}  
\usepackage{graphicx} 
\usepackage{natbib}  
\usepackage{caption} 
\usepackage{amsmath}
\usepackage{booktabs}
\usepackage{tikz}
\usetikzlibrary{positioning, arrows.meta}

\nocopyright

\title{Auditing Engagement Incentives in the Kidfluencer Ecosystem: A Multimodal Weak Supervision Approach\thanks{Preprint. Preliminary Work.}}

\author{
    Zijing Wei\textsuperscript{\rm 1},
    Chao Peter Yang\textsuperscript{\rm 2},
    Xuanjie Chen\textsuperscript{\rm 1}
}
\affiliations{
    \textsuperscript{\rm 1}Independent Contributor \\
    \textsuperscript{\rm 2}Duke University
}

\begin{document}

\maketitle

\begin{abstract}
The rapid rise of ``kidfluencers'' on YouTube has raised profound ethical concerns regarding child digital labor and exploitation. While emerging legislation attempts to regulate this ecosystem, empirical evidence on the relationship between child exploitation and engagement metrics remains scarce due to the challenge of operationalizing and scaling exploitation measurements. This study presents a multimodal AI audit of 5,051 videos across 79 kidfluencer channels, utilizing a weak supervision approach to detect exploitation signals without requiring large-scale manually labeled ground truth. We aggregate noisy labeling functions---including LLM-based classification of titles and GPT-4 Vision analysis of thumbnails and descriptions across six literature-grounded dimensions---to assign a probabilistic exploitation score to each video. A multi-annotator validation study ($N=107$ videos) combining independent human annotations demonstrates strong agreement with human judgment across six exploitation dimensions (macro-average F1 $= 0.911$) and high sensitivity for overall exploitation risk (recall $= 0.960$, overall F1 $= 0.793$).

Our findings reveal a highly significant \textbf{engagement premium associated with performative labor, emotional bait, and privacy violations}. Overall exploitation scores correlate significantly with view counts (Spearman $\rho = 0.229$, $p < 10^{-50}$). Mixed-effects regression, controlling for channel-level variation, demonstrates that a one-unit increase in exploitation score is associated with a 4.4$\times$ increase in views ($p < 0.001$). Within-channel analyses show that emotional bait and performative content are associated with median view boosts of $+65.6\%$ and $+56.0\%$ respectively (FDR-corrected $p<0.001$). These effects hold in robustness checks comparing videos published within the same year ($p=0.030$). Conversely, we find that explicit commercial content (product placement) does not benefit from this engagement premium ($-3.8\%$, n.s.), suggesting that the platform ecosystem rewards the commodification of the child's identity and labor rather than traditional advertising. These findings challenge policy frameworks focused solely on financial trusts, demonstrating that engagement metrics are systematically associated with the intensive, performative labor of children.
\end{abstract}

\section{Introduction}

The commercialization of childhood through digital platforms has created a lucrative ``kidfluencer'' economy in which children are featured in YouTube videos---unboxing toys, participating in challenges, performing scripted roleplay, and documenting their daily lives \citep{clark2025child}. This phenomenon has sparked intense debate regarding the ethics of child digital labor, the psychological impact of public exposure, and the blurring of lines between play and work, often termed ``playbour'' \citep{abidin2015micromicrocelebrity}.

Recent legislative efforts, such as the Illinois PA 103-0556 (2024) \citep{illinois2024}, aim to protect child creators by mandating financial trusts and limiting working hours. However, these regulations treat kidfluencing as a traditional labor market, often failing to address how platform ecosystems actively shape content creation \citep{unicef2025}. A fundamental question remains: \textbf{How do engagement metrics associate with specific dimensions of child exploitation?}

Previous research on the kidfluencer ecosystem has largely relied on qualitative content analysis or small-scale manual coding \citep{divon2025children,anderson2025growing}. While valuable, these approaches cannot scale to audit the massive volume of content generated. Conversely, purely computational approaches often struggle to operationalize complex, nuanced concepts like ``exploitation'' or ``performativity'' without expensive, large-scale human annotation. Furthermore, prior audits of platform algorithms \citep{huszar2022algorithmic,haroon2023auditing} have often focused on political content rather than child safety \citep{papadamou2020disturbed}.

This study bridges this gap by deploying a \textbf{multimodal weak supervision pipeline} to conduct a large-scale observational audit. We ground our conceptualization of exploitation risk in the UN Convention on the Rights of the Child (UNCRC), specifically drawing on Clark and Jno-Charles's \citep{clark2025child} five fundamental threats framework for kidfluencing. We operationalize these theoretical threats into six measurable content dimensions: performative labor, emotional bait, narrative conflict, challenge formats, commercial content, and privacy violations.

Crucially, because we rely on observational data via the YouTube API, we cannot directly observe the recommendation algorithm's internal scoring \citep{bouchaud2024auditing,rieder2018ranking}. Instead, we measure the \textbf{engagement premium}---the association between exploitative content dimensions and view counts---which serves as a proxy for the incentive structures shaping creator behavior.

Our core research questions are:
\begin{itemize}
    \item \textbf{RQ1:} Can a weak supervision framework effectively synthesize multimodal signals (text, LLM classifications, and computer vision) to measure kidfluencer exploitation at scale?
    \item \textbf{RQ2:} Are specific dimensions of exploitation (e.g., performative labor vs.\ commercial content) associated with higher view counts (an ``engagement premium'')?
    \item \textbf{RQ3:} Does this engagement premium persist \emph{within} channels and when controlling for video age, indicating a structural correlation rather than merely a channel-popularity effect?
\end{itemize}

\section{Related Work}

\subsection{The Kidfluencer Economy and Exploitation Frameworks}

The kidfluencer economy relies on the continuous documentation of children's private lives and their participation in scripted entertainment. Clark and Jno-Charles \citep{clark2025child} propose analyzing this phenomenon through the lens of the UNCRC, identifying five fundamental threats to children's rights in the kidfluencing context: (1) violation of consent, (2) violation of privacy, (3) economic exploitation (e.g., blurring work/leisure, making the child the product), (4) exposure to physical or psychological harm, and (5) suppression of authentic expression. Divon et al.\ \citep{divon2025children} further describe how children are transformed into ``concealed commodities'' through practices like transactional play. Because direct observation of rights violations (e.g., a child's true consent or compensation) is impossible via public metadata, computational audits must rely on proxy indicators of labor intensity and risk.

Other literature highlights the privacy risks of ``sharenting'' \citep{steinberg2017sharenting,kopecky2020sharenting}, where parents overshare children's lives online, potentially leading to emotional neglect \citep{keskin2023sharenting}. In extreme cases, such as the ``Elsagate'' phenomenon, platforms have struggled to moderate inappropriate content targeting toddlers \citep{papadamou2020disturbed,bridle2017something,tahir2019bringing}. Building on these frameworks, we operationalize exploitation not merely as overt abuse, but as the intensive, performative labor required to maintain engagement.

\subsection{Algorithmic Auditing and Engagement Metrics}

Algorithmic auditing investigates platform behavior without direct access to proprietary code, aiming to uncover systemic biases or perverse incentives \citep{sandvig2014auditing,metaxa2021auditing}. Previous studies have successfully audited YouTube and Twitter for political radicalization, algorithmic amplification of extreme content, and gender bias \citep{ribeiro2020auditing,huszar2022algorithmic,haroon2023auditing,hussein2020measuring}. These audits typically employ either active experimental designs, such as using sock puppet accounts to simulate user journeys \citep{habib2025auditing}, or large-scale observational analyses of engagement data \citep{bouchaud2024auditing}. 

However, the application of algorithmic auditing to child safety and the kidfluencer ecosystem remains limited. While platforms have implemented mechanisms like YouTube Kids to filter inappropriate content \citep{papadamou2020disturbed}, the incentive structures driving the creation of borderline exploitative content on the main platform are less understood. Because direct recommendation rates and internal algorithmic weights are hidden from external researchers, we must rely on observational proxies. Following established methodologies \citep{covington2016deep,zhou2010impact,rieder2018ranking}, we use engagement metrics (specifically, view counts) as a proxy for algorithmic reach and audience demand. By measuring the ``engagement premium'' (the positive association with specific content types), we can infer the economic incentives that likely shape creator behavior.

\subsection{Weak Supervision and LLM Content Analysis}

Traditional machine learning requires massive labeled datasets, which are difficult to obtain for subjective concepts. Weak supervision frameworks like Snorkel \citep{ratner2017snorkel,ratner2016data,bach2019snorkel} allow researchers to encode domain knowledge as noisy heuristic rules (Labeling Functions) to generate probabilistic labels \citep{johnson2022survey}. Recently, Large Language Models (LLMs) and Vision-Language Models (VLMs) have shown promise in zero-shot content moderation and annotation \citep{ma2023adapting,gilardi2023chatgpt,tornberg2024chatgpt}. We combine LLMs and VLMs within a weak supervision framework to scale our exploitation analysis across text and visual modalities.

\section{Methodology}

\subsection{Data Collection and Sampling}

We collected metadata for 58,965 videos from 79 family and kidfluencer YouTube channels using the YouTube Data API. Channels were selected based on prior literature and popular influencer lists, covering a spectrum of channel sizes and target audiences. The selected 79 channels represent a diverse cross-section of the kidfluencer ecosystem: 9 small channels ($<$50K median views), 31 medium (50K--500K), 26 large (500K--5M), and 13 very large ($>$5M median views), with an overall median of approximately 445,000 views per video. All channels are primarily English-language family vlog or kidfluencer channels featuring real children; animated channels were strictly excluded.

To manage computational costs while maintaining representativeness, we employed a stratified sampling strategy. For each of the 79 channels, we stratified videos into terciles based on view counts (high, medium, low) and randomly sampled up to 20 videos per tercile, resulting in a final stratified sample of 5,051 videos with valid view counts. We retrieved the title, description, thumbnail image, and publication date for each sampled video.

For the engagement analysis, we restricted the sample to 56 \textbf{kid-centric} channels ($n=4{,}208$ videos) by excluding 23 channels where the primary content creator is an adult (e.g., adult family vloggers who occasionally feature children). This ensures that the engagement premium analysis specifically captures incentives related to child-centered content. Within-channel pairwise comparisons further required sufficient within-channel variance in exploitation scores, yielding 47 channels with both exploitative and non-exploitative videos for the within-channel analysis.

\subsection{Exploitation Risk Dimensions}

Because actual exploitation (e.g., lack of consent, uncompensated labor) occurs off-camera, we cannot measure it directly. Instead, we operationalize Clark and Jno-Charles's \citep{clark2025child} five fundamental threats into six observable content dimensions that serve as proxy indicators of exploitation risk and labor intensity (Table~\ref{tab:mapping}).

\begin{table}[h!]
\centering
\small
\begin{tabular}{p{2.5cm} p{5cm}}
\toprule
\textbf{UNCRC Threat \citep{clark2025child}} & \textbf{Our Observable Dimension} \\
\midrule
Consent Violation & \textbf{Performative Labor:} Scripted/planned content suggesting directed labor rather than organic play. \\
Privacy Violation & \textbf{Privacy Violation:} Exposure of private, vulnerable, or medical moments. \\
Economic Exploitation & \textbf{Commercial Content:} Child used explicitly for product placement, making the child the product. \\
Exposure to Harm & \textbf{Challenge Format:} High-effort physical competitions.\\
& \textbf{Emotional Bait:} Exaggerated distress or emotional extremes used for clickbait. \\
Expression Suppression & \textbf{Narrative Conflict:} Manufactured drama replacing the child's authentic voice. \\
\bottomrule
\end{tabular}
\caption{Mapping theoretical UNCRC threats to our observable content dimensions.}
\label{tab:mapping}
\end{table}

\subsection{Multimodal Weak Supervision Pipeline}

\begin{figure}[t]
\centering
\resizebox{\columnwidth}{!}{%
\begin{tikzpicture}[node distance=0.6cm, >=stealth,
  box/.style={rectangle, draw=black!70, fill=blue!5, thick, minimum width=2.8cm, minimum height=0.55cm, align=center, font=\scriptsize},
  databox/.style={rectangle, draw=black!70, fill=green!5, thick, minimum width=2.8cm, minimum height=0.55cm, align=center, font=\scriptsize},
  resultbox/.style={rectangle, draw=black!70, fill=orange!8, thick, minimum width=2.8cm, minimum height=0.55cm, align=center, font=\scriptsize},
  arrow/.style={->, thick, black!70}]
  \node[databox] (data) {YouTube Data API\\58,965 videos, 79 channels};
  \node[box, below=of data] (sample) {Stratified Sampling\\5,051 videos};
  \node[box, below=0.8cm of sample] (llm) [xshift=-1.6cm] {33 Labeling Functions\\(LLM, VLM, Keywords,\\Metadata Heuristics)};
  \node[box, below=0.8cm of sample] (vlm) [xshift=1.6cm] {Snorkel Label Model\\(Learns LF accuracies\\and correlations)};
  \node[resultbox, below=0.8cm of llm, xshift=1.6cm] (agg) {Probabilistic Aggregation\\$\rightarrow$ Exploitation Score $\in [0,1]$};
  \node[resultbox, below=0.8cm of agg, xshift=-1.9cm] (val) {Human Validation\\$N{=}107$, Macro-F1{=}0.911};
  \node[resultbox, below=0.8cm of agg, xshift=1.9cm] (analysis) {Engagement Analysis\\Mixed Effects + FDR};
  \draw[arrow] (data) -- (sample);
  \draw[arrow] (sample) -- (llm);
  \draw[arrow] (sample) -- (vlm);
  \draw[arrow] (llm) -- (agg);
  \draw[arrow] (vlm) -- (agg);
  \draw[arrow] (agg) -- (val);
  \draw[arrow] (agg) -- (analysis);
\end{tikzpicture}%
}
\caption{The multimodal weak supervision pipeline. LLM text and VLM multimodal scores are aggregated into a probabilistic exploitation score, validated against human annotations.}
\label{fig:pipeline}
\end{figure}
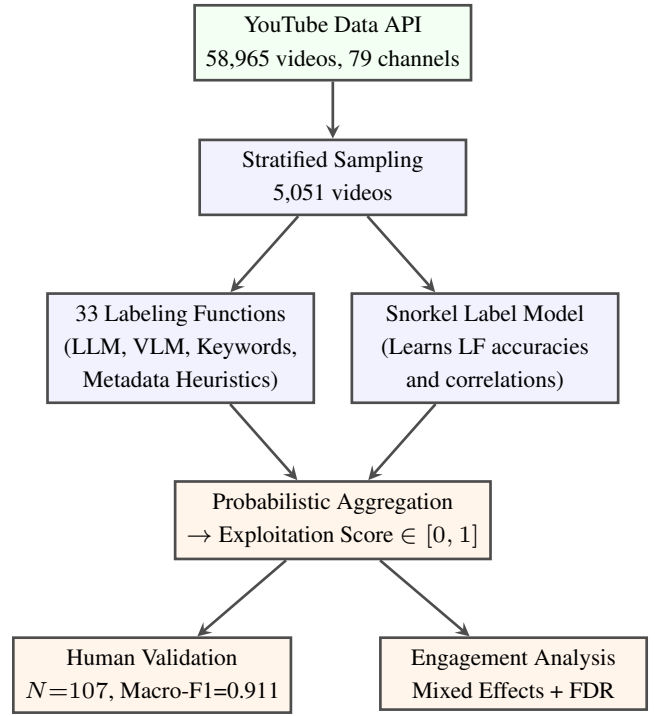

We implemented a multimodal weak supervision pipeline (Figure~\ref{fig:pipeline}) using the Snorkel framework \citep{ratner2017snorkel}. Rather than relying on a single model or manual weighting, we designed 33 heterogeneous Labeling Functions (LFs) across the six dimensions. These LFs capture diverse signals from the metadata:
\begin{itemize}
    \item \textbf{LLM and VLM Zero-Shot Classifiers:} We deployed GPT-4.1-mini (text-only, analyzing titles) and GPT-4.1-mini Vision (analyzing titles, thumbnails, and descriptions) as high-capacity LFs.
    \item \textbf{Keyword and Pattern Heuristics:} Regular expressions detecting explicit clickbait patterns (e.g., ALL CAPS, excessive punctuation, ``GONE WRONG''), emotional triggers (e.g., specific emojis, words like ``crying'' or ``scared''), and commercial tags (e.g., ``\#ad'', ``sponsored'').
    \item \textbf{Metadata Heuristics:} Rules based on video duration (e.g., videos over 30 minutes are more likely to be extended challenges) and description structures.
\end{itemize}

For each dimension, a Snorkel Label Model was trained to automatically estimate the unknown accuracies and correlations of these LFs without ground truth data. The Label Model aggregates the noisy, conflicting signals from the 33 LFs into a single probabilistic exploitation score ($\in [0, 1]$) for each video. For example, in the performative labor dimension, the Label Model learned to assign high weights to the VLM's high-confidence predictions ($w=0.962$) and specific organic-content keyword rules ($w=0.977$), while down-weighting noisier heuristics. This data-driven aggregation ensures that our final exploitation scores are robust to the failure modes of any single LF or model.

\subsubsection{Human Validation and Annotation Protocol.}

To rigorously validate the multimodal pipeline, we conducted a multi-annotator validation study. Three independent annotators manually labeled a stratified random sample of 107 unique videos from kid-centric channels. Annotators were trained on the Clark and Jno-Charles \citep{clark2025child} five fundamental threats framework and assessed the overall exploitation risk of each video by reviewing the title, thumbnail, and at least 30 seconds of video content. The annotation protocol instructed coders to identify performative labor when the child was clearly engaging in scripted or planned activities solely for the camera (e.g., skits, structured games), as opposed to organic documentation of daily life (e.g., vlogs of shopping or moving). Emotional bait was coded when the title or thumbnail explicitly manufactured exaggerated emotional states (e.g., shock, fear, guilt) to attract clicks.

\begin{table}[t]
\centering
\small
\begin{tabular}{lrrrr}
\toprule
\textbf{Dimension} & \textbf{Prev} & \textbf{Prec} & \textbf{Rec} & \textbf{F1} \\
\midrule
Performative Labor & 69.8\% & 0.923 & 0.973 & 0.947 \\
Emotional Bait & 85.0\% & 0.944 & 1.000 & 0.971 \\
Narrative Conflict & 59.5\% & 1.000 & 0.955 & 0.977 \\
Challenge Format & 36.7\% & 1.000 & 0.944 & 0.971 \\
Commercial Content & 51.0\% & 0.641 & 1.000 & 0.781 \\
Privacy Violation & 29.5\% & 1.000 & 0.692 & 0.818 \\
\midrule
\textbf{Macro-Average} & & & & \textbf{0.911} \\
\bottomrule
\end{tabular}
\caption{Per-dimension validation of the multimodal pipeline against consensus human annotations ($N=53$ videos with per-dimension labels).}
\label{tab:validation_dims}
\end{table}

In total, we compiled 191 annotation entries across 107 unique videos that matched our kid-centric Snorkel dataset. For the 53 videos where detailed per-dimension labels were available, we computed consensus labels via majority vote. Table~\ref{tab:validation_dims} reports the model's performance on these individual dimensions. The pipeline demonstrated exceptionally strong capability in identifying specific exploitation signals, achieving a macro-average F1 score of $0.911$. Notably, the model achieved near-perfect F1 scores ($>0.94$) for performative labor, emotional bait, narrative conflict, and challenge formats. Commercial content exhibited lower precision ($0.641$) despite perfect recall; this occurred because the model detected subtle commercial signals (e.g., brand mentions in descriptions) that human annotators did not flag unless explicit sponsorship was present. Similarly, privacy violations showed lower recall ($0.692$) as the model missed a few cases where implicit privacy issues were not fully captured by metadata and visual heuristics alone.

For the overall binary exploitation risk, we evaluated the pipeline against the full set of 107 unique annotated videos. Using an exploitation score threshold of $0.5$, the model achieved an overall accuracy of $0.766$, precision of $0.676$, recall of $0.960$, and an F1 score of $0.793$. The model's primary error mode was false positives (moderate precision), reflecting a deliberate design choice to err on the side of flagging potential exploitation (high recall), consistent with the audit's protective intent. Human annotators generally applied a higher, more conservative threshold for determining whether a video was "overall exploitative," even when 1-2 mild exploitation signals were present.

The pipeline computed a combined probability score for each dimension, and an \textbf{Overall Exploitation Score} $\in [0, 1]$ for each video.

\section{Results}

\subsection{Pipeline Performance and Dimension Prevalence}

The multimodal pipeline successfully processed 4,208 videos from 56 kid-centric channels. Overall, 26.5\% of the sampled videos ($n=1,114$) were classified as exploitative (score $\ge 0.7$).

\begin{table*}[t]
\centering
\begin{tabular}{lrrrrr}
\toprule
\textbf{Exploitation Dimension} & \textbf{Prevalence} & \textbf{Mean Log-Premium} & \textbf{Cohen's $d$} & \textbf{FDR $p$-value} & \textbf{Mixed-Effects $\beta$} \\
\midrule
Performative Labor & 69.5\% & $+0.193$ & $+0.524$ & $<0.001^{***}$ & $0.091^{***}$ \\
Emotional Bait & 56.2\% & $+0.219$ & $+0.751$ & $<0.001^{***}$ & $0.264^{***}$ \\
Narrative Conflict & 34.9\% & $+0.145$ & $+0.511$ & $<0.001^{***}$ & $0.153^{***}$ \\
Challenge Format & 32.4\% & $+0.082$ & $+0.320$ & $0.081$ (n.s.) & $0.003$ (n.s.) \\
Commercial Content & 41.4\% & $-0.017$ & $-0.073$ & $0.516$ (n.s.) & $0.042$ (n.s.) \\
Privacy Violation & 22.9\% & $+0.147$ & $+0.650$ & $<0.001^{***}$ & $0.226^{***}$ \\
\midrule
\textbf{Overall Exploitation} & 26.5\% & $+0.123$ & $+0.582$ & $<0.001^{***}$ & $0.647^{***}$ \\
\bottomrule
\end{tabular}
\caption{Summary of engagement premiums by exploitation dimension. Mean log-premium represents the within-channel difference in $\log_{10}(\text{views})$. Mixed-effects $\beta$ represents the coefficient in the joint model controlling for all dimensions simultaneously.}
\label{tab:main_results}
\end{table*}

\begin{figure}[h!]
\centering
\includegraphics[width=\columnwidth]{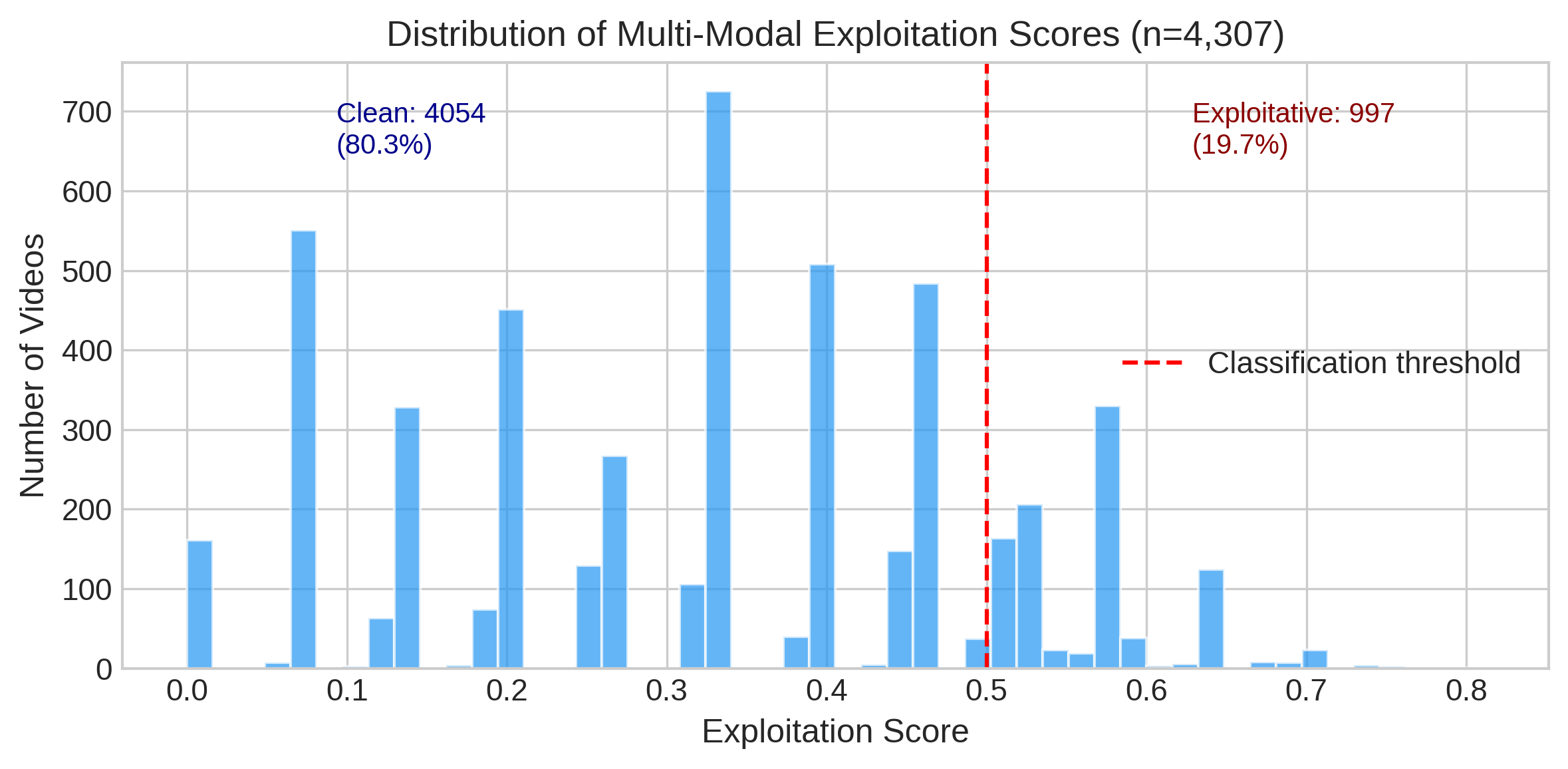}
\caption{Distribution of the probabilistic exploitation score generated by the multimodal weak supervision model.}
\label{fig:score_dist}
\end{figure}

Performative labor was the most prevalent dimension (detected in 69.5\% of videos), followed by emotional bait (56.2\%), commercial content (41.4\%), narrative conflict (34.9\%), challenge formats (32.4\%), and privacy violations (22.9\%). Overall, using an aggregated probability threshold of 0.7, 26.5\% of videos in the kid-centric dataset were classified as high exploitation risk.

\subsection{The Engagement Premium (RQ2 \& RQ3)}

We found a highly significant positive correlation between a video's continuous exploitation risk score and its view count (Spearman $\rho = 0.229$, $p < 10^{-50}$).

To determine if this association is a structural platform correlation rather than merely an artifact of highly exploitative channels being more popular overall, we conducted two primary analyses: mixed-effects regression and within-channel pairwise comparisons.

\subsubsection{Mixed-Effects Regression.}

We fit a mixed-effects linear regression model predicting $\log_{10}(\text{views})$ from the overall exploitation score, including random intercepts for each of the 56 kid-centric channels to control for baseline channel popularity. The model specification is:
\[
\log_{10}(\text{views}_{ij}) = \beta_0 + \beta_1 \cdot \text{ExploitScore}_{ij} + u_j + \epsilon_{ij}
\]
where $u_j \sim N(0, \sigma^2_u)$ represents the channel-level random intercept for channel $j$, and $\epsilon_{ij}$ is the residual error. This formulation ensures that the estimated effect of exploitation score reflects \emph{within-channel} variation rather than between-channel differences in baseline popularity.

The model revealed a massive, highly significant effect: a one-unit increase in the exploitation risk score is associated with a $0.647$ increase in $\log_{10}(\text{views})$ ($\beta = 0.647$, $SE = 0.059$, $z = 10.91$, $p < 0.001$). This translates to approximately a \textbf{4.4$\times$ increase in raw view counts} ($10^{0.647} \approx 4.4$). The random intercept variance was substantial ($\sigma^2_u = 0.741$), confirming that channels vary considerably in baseline popularity and justifying the multilevel approach. The intraclass correlation coefficient (ICC $= 0.738$) indicates that 73.8\% of the variance in $\log_{10}(\text{views})$ is attributable to between-channel differences, underscoring the importance of controlling for channel identity.

We then fit a second mixed-effects model using the continuous probabilities of the six individual dimensions as fixed effects. When controlling for all dimensions simultaneously, \textbf{emotional bait} ($\beta = 0.264$, $p < 0.001$), \textbf{privacy violations} ($\beta = 0.226$, $p < 0.001$), \textbf{narrative conflict} ($\beta = 0.153$, $p < 0.001$), and \textbf{performative labor} ($\beta = 0.091$, $p < 0.001$) remained highly significant predictors of increased viewership. Challenge formats ($\beta = 0.003$, $p = 0.91$) and commercial content ($\beta = 0.042$, $p = 0.24$) were not statistically significant in the joint model, suggesting their effects may be mediated by correlation with the primary dimensions.

\subsubsection{Within-Channel Pairwise Comparisons.}

For each dimension, we compared the median views of videos exhibiting that dimension against videos from the \emph{same channel} lacking it. To account for multiple comparisons, we applied False Discovery Rate (FDR) correction using the Benjamini-Hochberg procedure.

\begin{figure}[h!]
\centering
\includegraphics[width=\columnwidth]{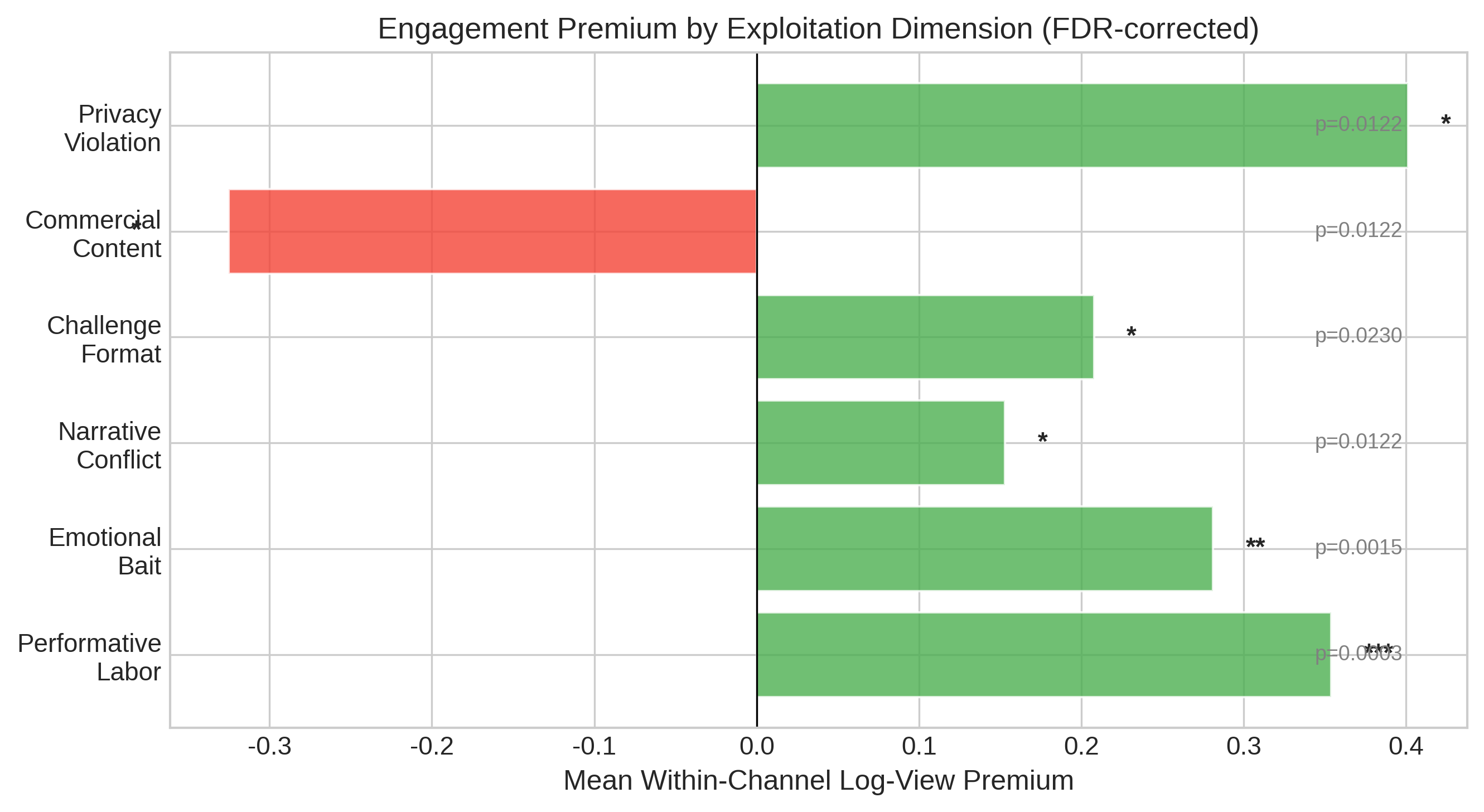}
\caption{Mean within-channel view boost by exploitation dimension (FDR-corrected). Error bars represent 95\% confidence intervals.}
\label{fig:premiums}
\end{figure}

The results reveal a clear engagement premium structure across channels with sufficient data:
\begin{itemize}
    \item \textbf{Emotional Bait} is associated with the largest premium, generating $+65.6\%$ more views than non-emotional videos on the same channel (Cohen's $d = 0.751$, FDR $p<0.001$).
    \item \textbf{Performative Labor} is associated with a $+56.0\%$ view premium ($d = 0.524$, FDR $p<0.001$).
    \item \textbf{Privacy Violations} and \textbf{Narrative Conflict} are associated with $+40.3\%$ ($d = 0.650$, FDR $p<0.001$) and $+39.7\%$ ($d = 0.511$, FDR $p<0.001$) premiums respectively.
    \item \textbf{Challenge Formats} showed a smaller, non-significant positive premium ($+20.9\%$, FDR $p=0.081$).
\end{itemize}

Crucially, \textbf{Commercial Content} exhibited a slight \textbf{negative} premium ($-3.8\%$, FDR $p=0.516$). While not statistically significant after FDR correction, the directionality suggests that videos featuring explicit product placements do not benefit from the massive engagement boosts seen in emotional and performative exploitation dimensions.

\begin{figure}[h!]
\centering
\includegraphics[width=\columnwidth]{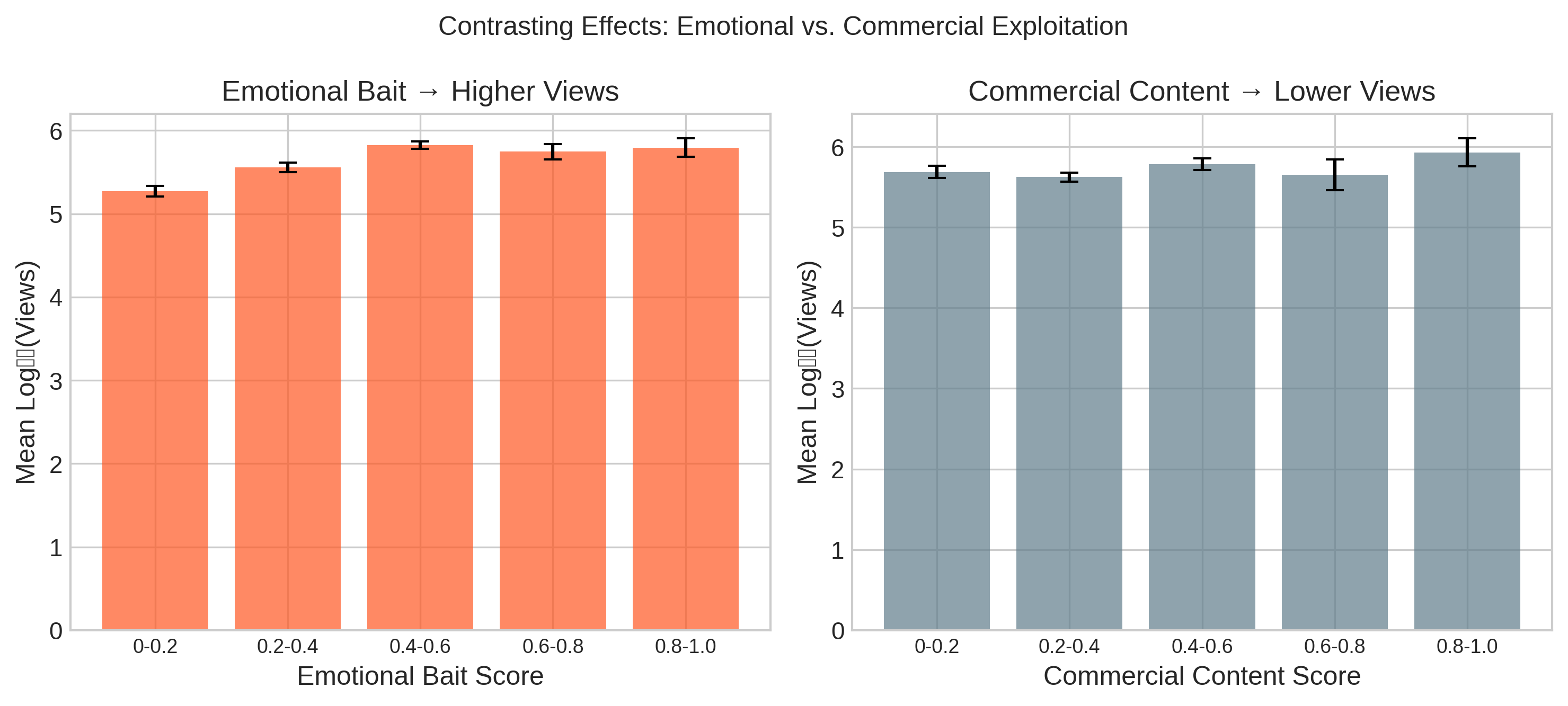}
\caption{Contrasting effects: Emotional exploitation dimensions show positive engagement premiums while commercial content shows a negative premium.}
\label{fig:contrast}
\end{figure}

\subsection{Robustness Checks}
\label{sec:robustness}

To ensure our findings were not merely artifacts of video age (older videos accumulating more views), we conducted a \textbf{same-year within-channel comparison} on the subset of videos with valid ISO-8601 publication dates ($N=907$). By matching exploitative and non-exploitative videos published by the same channel in the same year (23 channel-year groups), we found the engagement premium holds robustly: high-exploitation videos received $+44.4\%$ more views than their same-year, same-channel counterparts (Wilcoxon $p=0.030$).

\begin{figure}[h!]
\centering
\includegraphics[width=\columnwidth]{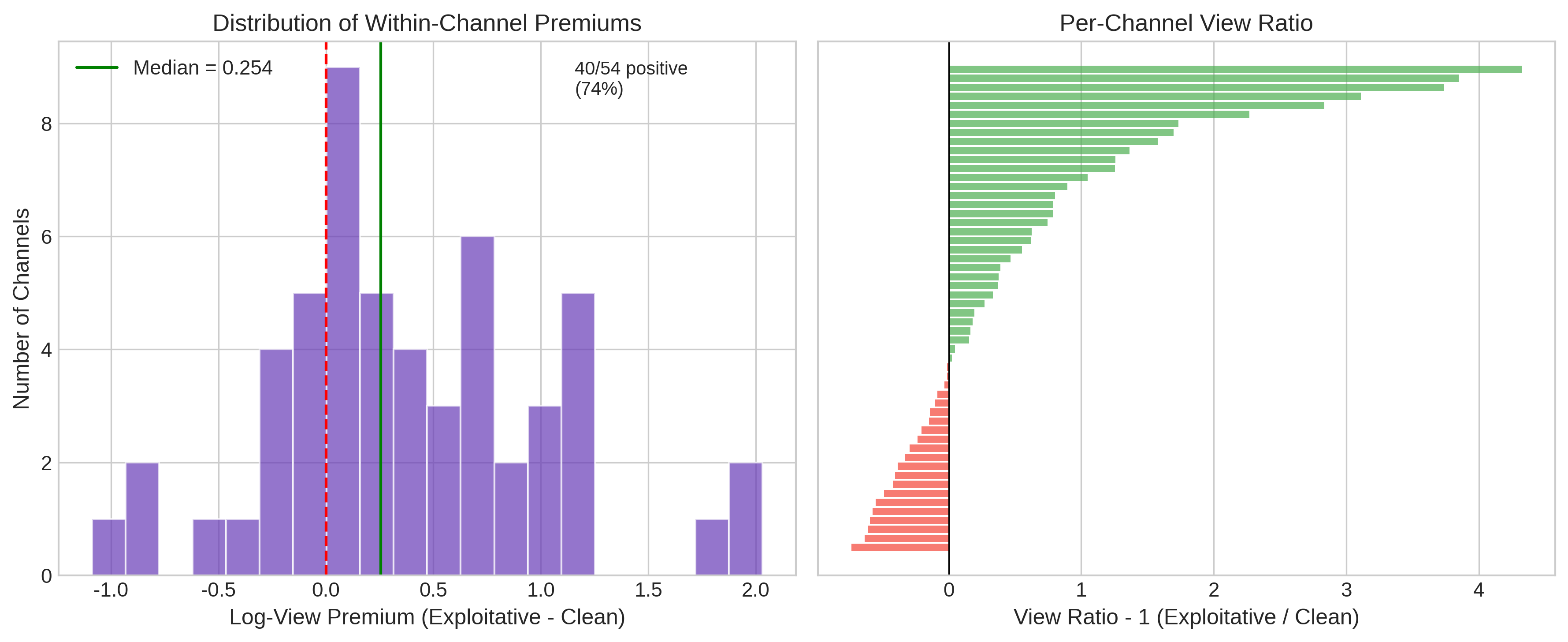}
\caption{Distribution of overall within-channel premiums. 76.6\% of channels exhibit a positive premium for exploitative content.}
\label{fig:channel_dist}
\end{figure}

Additionally, we verified the robustness of our results by analyzing the overall exploitation flag at the channel level. Across 47 channels with sufficient variance, 36 channels (76.6\%) exhibited a positive engagement premium for high-exploitation videos, yielding an average within-channel boost of $+32.8\%$ (Wilcoxon $p=0.001$). This confirms that the performativity premium is a widespread structural phenomenon, not driven by a few outlier channels.

\section{Discussion}

\subsection{The ``Performativity Premium'' vs.\ The Commercial Non-Effect}

Our findings provide large-scale empirical evidence for a ``performativity premium'' in the kidfluencer ecosystem. Engagement metrics are systematically associated with content features that proxy for intensive, performative labor (scripted conflict, emotional bait) over organic family documentation. This premium creates a structural incentive gradient where creators who push boundaries, manufacture drama, and expose children's emotional vulnerabilities are systematically rewarded with higher attention in a highly competitive market. Crucially, while our metrics capture content features rather than direct UNCRC violations, the high prevalence of these features (e.g., 69.5\% performative labor) suggests that the labor intensity required to succeed on the platform is structurally high.

Interestingly, traditional commercial content (e.g., explicit toy unboxing or product placement) does not benefit from this engagement premium ($-3.8\%$, n.s.). This suggests a fundamental shift in the kidfluencer economy: the most successful strategy for driving massive viewership is not to use the child to sell a physical toy, but to make the child's labor, emotions, and manufactured drama the product itself. This aligns with Divon et al.'s \citep{divon2025children} concept of ``transactional childhood.'' Audiences appear less responsive to overt advertising but are heavily drawn to the commodification of the child's identity. Furthermore, the lack of an engagement premium for commercial content may partially reflect platform-level interventions; following the FTC's COPPA settlement, YouTube implemented stricter policies on ``made for kids'' content, potentially reducing the algorithmic reach or monetization potential of explicit advertising aimed at minors. 

\subsection{Qualitative Nuances: Vlog vs.\ Staged Drama}

During our human validation process, we observed that distinguishing between organic documentation (vlogging) and performative labor (staged drama) is highly nuanced. Some channels operate with a fundamentally exploitative premise (e.g., daily scripted pranks), while others primarily document family life but occasionally use exaggerated thumbnails (emotional bait) to drive clicks. Our multimodal pipeline's ability to output a continuous probability score ($\in [0, 1]$) rather than a binary label effectively captures this spectrum of exploitation. Videos with clickbait thumbnails but organic content received moderate scores, while highly scripted, high-conflict videos received the highest scores. This continuous measurement highlights that exploitation in the kidfluencer space is rarely binary; it exists on a spectrum driven by the constant pressure to optimize for engagement.

\subsection{Case Study: The Industrialized Kidfluencer Network}

To illustrate the structural dynamics our pipeline captures, we briefly describe the operational characteristics of one of the most-subscribed children's channel networks (anonymized). This network operates 13 YouTube channels across multiple languages, collectively reaching over 300 million subscribers. A 50-person professional production team manages daily content output, including scripted narratives, custom animations, elaborate physical sets, and post-production editing indistinguishable from broadcast television. Each child in the family maintains independent accounts across YouTube, TikTok, Instagram, and Facebook, with near-daily uploads on each platform. Thumbnails follow a systematic design strategy: saturated colors, children's faces in exaggerated emotional expressions, and text overlays engineered for maximum click-through rates.

Crucially, while this operation presents itself as organic family content, the scale of production (daily uploads across 13 channels with professional crews) represents a labor intensity that far exceeds what any individual family could sustain without treating content creation as a full-time commercial enterprise. In our framework, such channels consistently score high on performative labor and commercial content dimensions---not because any single video is necessarily harmful, but because the \emph{systemic pattern} reveals an industrial operation built around children's continuous on-camera presence.

\subsection{Proxy Indicators and the Limits of Observational Measurement}

A critical epistemological caveat must be acknowledged: our exploitation score measures the \emph{intensity of content features associated with exploitation risk}, not exploitation itself. A high score indicates that a video exhibits characteristics---scripted performativity, emotional manipulation in metadata, commercial framing---that prior literature identifies as risk factors \citep{clark2025child,divon2025children}. However, a child may enthusiastically participate in a challenge video without experiencing harm, and conversely, exploitation may occur in videos that appear benign from metadata alone.

We therefore frame our findings not as evidence that specific children are being exploited, but as evidence that \textbf{platform incentive structures systematically reward content features associated with exploitation risk}. This framing strengthens rather than weakens our policy argument: regardless of whether any individual family is exploiting their children, the engagement premium we document creates structural pressure toward increasingly intensive child labor. The policy concern is not any single video, but the cumulative incentive gradient that shapes creator behavior across the ecosystem.

\subsection{Policy Implications}

Our findings have direct implications for the ongoing policy debate surrounding child digital labor. Current legislative frameworks, such as the Illinois PA 103-0556 and the proposed federal KIDS Act (S. 1409), focus primarily on financial compensation---mandating trust funds and limiting working hours. While these protections are necessary, our results suggest they are insufficient. The engagement premium we document operates independently of direct monetary compensation: a child's performative labor is strongly associated with views (and thus advertising revenue) regardless of whether the child receives a share of that income.

Platform-level interventions may be more effective than purely legislative approaches. Our finding that commercial content does not benefit from the engagement premium suggests that YouTube's existing policies on ``made for kids'' content may already be partially effective at reducing the incentive for overt commercialization. Similar mechanisms could be extended to address the performativity premium: for instance, platforms could reduce algorithmic amplification of content flagged as involving intensive child labor, or implement friction mechanisms (e.g., mandatory cooling-off periods) for channels that consistently produce high-exploitation content. However, such interventions must be carefully designed to avoid penalizing legitimate family documentation or cultural expression.

\subsection{Methodological Contributions}

This study demonstrates the efficacy of multimodal weak supervision for large-scale observational auditing. By combining LLM text analysis and VLM visual analysis, we successfully scaled the operationalization of complex ethical concepts without requiring massive manual ground truth. Our multi-annotator validation (overall F1 $= 0.793$, recall $= 0.960$, macro-average per-dimension F1 $= 0.911$) demonstrates that the pipeline reliably identifies specific exploitation risk indicators in a manner consistent with trained human judgment. The approach is particularly well-suited to domains where ground truth is inherently subjective and expensive to obtain at scale, such as hate speech detection, misinformation assessment, and content safety evaluation.

\subsection{Limitations and Future Work}

This study has several limitations. First, while we conducted a multi-annotator validation ($N=107$), we observed that different annotators exhibited varying thresholds for defining exploitation risk, highlighting the inherent subjectivity of the concept. For example, a more conservative annotator often labeled family game night videos as low-risk, whereas the primary annotator flagged them as performative labor. While inter-rater agreement on overlapping samples was perfect (Cohen's $\kappa=1.0$), this was based on a small overlap set ($n=3$); future work should expand the overlap set to systematically study these subjective disagreements. Furthermore, our dimensions are proxy indicators of exploitation risk, not definitive proof of UNCRC violations. A child may happily participate in a challenge video without being exploited, but the systemic prevalence of such content indicates a high baseline of labor intensity.

Second, our analysis relies on observational data. We measure \emph{engagement metrics} (views), which are proxies for algorithmic reach, but we cannot claim a direct causal link to YouTube's internal recommendation weights \citep{rieder2018ranking}. The engagement premium we observe may reflect audience preferences, algorithmic amplification, or both.

Third, our VLM classification primarily analyzed video titles, thumbnails, and descriptions; future work should incorporate full video transcripts and duration data to capture intra-video dynamics. A video's thumbnail may suggest emotional bait while the actual content is benign, or conversely, a neutral thumbnail may mask exploitative content within the video itself. Incorporating temporal signals (e.g., video duration, upload frequency) could also reveal patterns of labor intensity that metadata alone cannot capture.

Fourth, our sample is limited to English-language channels based in North America and the UK. The kidfluencer economy operates globally, with significant markets in South Korea, Brazil, India, and Southeast Asia. Cross-cultural analyses may reveal different exploitation patterns and engagement dynamics shaped by distinct regulatory environments, cultural norms around childhood, and platform-specific affordances (e.g., YouTube Shorts vs.\ long-form content).

Finally, our study captures a cross-sectional snapshot. Longitudinal analyses tracking how individual channels evolve their content strategies in response to engagement feedback would provide stronger evidence for the causal mechanisms underlying the performativity premium.

\section{Ethics and Positionality Statement}

\subsection{Ethical Considerations}
This research utilizes publicly available, observational data from YouTube via the official API. The study was deemed IRB-exempt by our institutional review board as it involves no direct interaction with human subjects or minors, and analyzes public figures participating in the digital economy. However, we recognize that the subjects of these videos are children who cannot legally consent to their digital footprint. To minimize potential harm, we adhered to strict data minimization principles. We did not download or store video files; we only retrieved metadata, thumbnails, and descriptions necessary for the analysis. Furthermore, we report all findings as aggregate statistics or anonymized trends. When discussing specific examples or qualitative findings, we intentionally obscure identifying details to prevent further amplification or privacy violations of the children involved. Our goal is to audit the systemic incentives of the platform ecosystem, not to target or shame individual families.

\subsection{Positionality Statement}
The authors approach this research from the perspective of computational social scientists and algorithmic auditors based in North America. We acknowledge that our operationalization of ``exploitation'' is grounded in Western frameworks of child rights (e.g., the UNCRC) and North American legislative contexts (e.g., the Illinois PA 103-0556). The kidfluencer economy operates globally, and cultural norms regarding child labor, family documentation, and privacy vary significantly. Our analysis of English-language channels may not capture the nuances of exploitation in other cultural contexts. We also recognize the inherent subjectivity in labeling content as ``exploitative.'' By employing a weak supervision approach and providing transparent validation metrics, we aim to make our assumptions explicit and open to critique, rather than presenting our pipeline's outputs as absolute ground truth.

\section{Conclusion}

As the kidfluencer economy matures, regulatory focus must expand beyond financial compensation to address the structural forces shaping content creation. Our multimodal audit of 4,208 videos across 56 kid-centric channels demonstrates that engagement metrics are significantly associated with content featuring child performative labor, privacy violations, and emotional bait, while overt commercial content does not benefit from this premium. By highlighting this ``performativity premium,'' this study underscores the need for platform-level interventions that disincentivize the commodification of child labor and stress.

\bibliography{references}

\appendix

\section{Labeling Function Inventory}
\label{app:lfs}

Table~\ref{tab:lf_inventory} provides the complete list of 33 labeling functions (LFs) used in our weak supervision pipeline. Each LF is trained independently per dimension using Snorkel's label model, which learns the accuracy and correlation structure of the LFs without requiring ground truth labels.

\begin{table*}[h!]
\centering
\small
\begin{tabular}{p{1.8cm} l c c p{5.5cm}}
\toprule
\textbf{Dim.} & \textbf{LF Name} & \textbf{Type} & \textbf{Cov.} & \textbf{Description} \\
\midrule
\multirow{8}{=}{\centering Perf. Labor}
& lf\_llm\_performative & LLM & 1.000 & GPT-4.1-nano classifies title as performative (scripted/staged) vs.\ organic \\
& lf\_vlm\_performative\_high & VLM & 0.780 & GPT-4.1-mini Vision scores $\ge 0.7$ for performative labor \\
& lf\_vlm\_performative\_med & VLM & 0.932 & GPT-4.1-mini Vision scores $\ge 0.4$ for performative labor \\
& lf\_scripted\_keywords & Rule & 0.253 & Title contains scripted content keywords (skit, challenge, prank, etc.) \\
& lf\_organic\_keywords & Rule & 0.076 & Title contains organic family keywords (birthday, vacation) $\rightarrow$ NOT exploit \\
& lf\_roleplay\_title & Rule & 0.031 & Title contains roleplay/acting keywords \\
& lf\_desc\_scripted & Rule & 0.408 & Description contains scripted production indicators \\
& lf\_duration\_long & Meta & 0.035 & Video duration $>$ 30 minutes (long-form scripted content) \\
\midrule
\multirow{6}{=}{\centering Emot. Bait}
& lf\_llm\_emotional & LLM & 0.100 & GPT-4.1-nano classifies title as emotional bait \\
& lf\_vlm\_emotional & VLM & 0.625 & GPT-4.1-mini Vision scores emotional bait from thumbnail+title \\
& lf\_emotional\_caps & Rule & 0.179 & Title has $>60\%$ uppercase characters (ALL CAPS clickbait) \\
& lf\_emotional\_punct & Rule & 0.063 & Title contains $\ge 3$ exclamation marks \\
& lf\_emotional\_keywords & Rule & 0.018 & Title contains emotional distress words (crying, meltdown, etc.) \\
& lf\_emotional\_emojis & Rule & 0.001 & Title contains emotional emojis (crying face, etc.) \\
\midrule
\multirow{5}{=}{\centering Narr. Confl.}
& lf\_llm\_narrative & LLM & 0.091 & GPT-4.1-nano classifies title as narrative conflict \\
& lf\_vlm\_narrative & VLM & 0.753 & GPT-4.1-mini Vision scores narrative conflict \\
& lf\_conflict\_keywords & Rule & 0.056 & Title contains conflict keywords (fight, stolen, arrested, etc.) \\
& lf\_prank\_gone\_wrong & Rule & 0.016 & Title contains ``prank'' or ``gone wrong'' patterns \\
& lf\_clickbait\_patterns & Rule & 0.004 & Title matches clickbait structural patterns \\
\midrule
\multirow{5}{=}{\centering Challenge}
& lf\_llm\_challenge & LLM & 0.175 & GPT-4.1-nano classifies title as challenge format \\
& lf\_vlm\_challenge & VLM & 0.903 & GPT-4.1-mini Vision scores challenge format \\
& lf\_challenge\_keywords & Rule & 0.136 & Title contains challenge keywords (24 hours, last to, vs, etc.) \\
& lf\_competition\_format & Rule & 0.051 & Title matches competition format patterns \\
& lf\_duration\_challenge & Meta & 0.003 & Video duration matches typical challenge length \\
\midrule
\multirow{5}{=}{\centering Commercial}
& lf\_llm\_commercial & LLM & 0.032 & GPT-4.1-nano classifies title as commercial content \\
& lf\_vlm\_commercial & VLM & 0.715 & GPT-4.1-mini Vision scores commercial signals from thumbnail \\
& lf\_commercial\_title & Rule & 0.136 & Title contains brand names or product references \\
& lf\_commercial\_desc & Rule & 0.129 & Description contains affiliate links, discount codes, or \#ad \\
& lf\_commercial\_tags & Rule & 0.345 & Video tags contain commercial/brand indicators \\
\midrule
\multirow{4}{=}{\centering Privacy}
& lf\_llm\_privacy & LLM & 0.027 & GPT-4.1-nano classifies title as privacy violation \\
& lf\_vlm\_privacy & VLM & 0.818 & GPT-4.1-mini Vision scores privacy concern from thumbnail \\
& lf\_privacy\_keywords & Rule & 0.034 & Title contains privacy-sensitive keywords (bath, potty, etc.) \\
& lf\_privacy\_medical\_desc & Rule & 0.013 & Description mentions medical/health details of child \\
\bottomrule
\end{tabular}
\caption{Complete inventory of 33 labeling functions. Type: LLM = GPT-4.1-nano text classification; VLM = GPT-4.1-mini Vision multimodal analysis; Rule = keyword/regex heuristic; Meta = metadata-based (duration, tags). Coverage = proportion of videos receiving a non-abstain vote.}
\label{tab:lf_inventory}
\end{table*}

\section{Snorkel Label Model Diagnostics}
\label{app:diagnostics}

Table~\ref{tab:diagnostics} reports the standard diagnostic metrics for our Snorkel label model, computed per dimension.

\begin{table*}[h!]
\centering
\begin{tabular}{lrrrr}
\toprule
\textbf{Dimension} & \textbf{\# LFs} & \textbf{Abstain \%} & \textbf{Mean Cov.} & \textbf{Mean Conf.} \\
\midrule
Performative Labor & 8 & 0.0\% & 0.439 & 0.243 \\
Emotional Bait & 6 & 29.2\% & 0.164 & 0.009 \\
Narrative Conflict & 5 & 23.1\% & 0.184 & 0.002 \\
Challenge Format & 5 & 8.7\% & 0.253 & 0.008 \\
Commercial Content & 5 & 14.7\% & 0.271 & 0.096 \\
Privacy Violation & 4 & 16.2\% & 0.223 & 0.012 \\
\midrule
\textbf{Aggregate (33 LFs)} & 33 & --- & 0.271 & 0.078 \\
\bottomrule
\end{tabular}
\caption{Snorkel label model diagnostics per dimension. Abstain \% = proportion of videos where no LF votes (label model uses prior). Mean Cov.\ = average LF coverage. Mean Conf.\ = average pairwise conflict rate among LFs.}
\label{tab:diagnostics}
\end{table*}

\section{LLM and VLM Prompts}
\label{app:prompts}

\subsection{LLM Title Classification Prompt (GPT-4.1-nano)}

The following system prompt is used for text-only classification of video titles:

\begin{quote}
\small
\texttt{You are an expert in children's media and child exploitation research. You will classify YouTube video titles from kidfluencer/family vlog channels along 5 dimensions. For each title, output a JSON object with these fields:}

\texttt{- ``performative'': 1 if the child is clearly performing/working FOR the video (challenges, roleplay, scripted skits, dance routines, unboxing, reviews, pranks, games designed for content). 0 if organic/natural (birthday, vacation, daily life that would happen without a camera). -1 if ambiguous.}

\texttt{- ``emotional\_bait'': 1 if the title uses exaggerated emotional language or manufactured drama to attract clicks. This includes: ALL CAPS shouting, excessive punctuation, fake emergencies, exaggerated reactions, manufactured urgency. 0 if calm/descriptive.}

\texttt{- ``narrative\_conflict'': 1 if the title implies interpersonal conflict, mystery, or dramatic tension (theft, confrontation, betrayal, punishment, secrets revealed). 0 if no narrative tension.}

\texttt{- ``challenge\_format'': 1 if the title indicates a challenge, competition, or game format (``24 HOURS'', ``LAST TO LEAVE'', ``VS'', dares, contests). 0 if not.}

\texttt{- ``commercial\_content'': 1 if the title references specific brands, products, stores, or commercial activities. 0 if no brand/product reference.}
\end{quote}

\subsection{VLM Multimodal Prompt (GPT-4.1-mini Vision)}

The following prompt is sent with the video thumbnail (low-detail mode), title, and first 300 characters of the description:

\begin{quote}
\small
\texttt{You are analyzing a YouTube video thumbnail and title to assess potential child exploitation signals. The video is from a family/kid YouTube channel.}

\texttt{Analyze the thumbnail image and title together. For each dimension below, provide a score from 0.0 to 1.0:}

\texttt{1. performative\_labor (0.0--1.0): Is the child performing scripted/staged content? Score HIGH if the activity would NOT happen without the camera.}

\texttt{2. emotional\_bait (0.0--1.0): Does the thumbnail/title use exaggerated emotions to attract clicks? Look for: children with mouths wide open in shock, crying faces, ALL CAPS.}

\texttt{3. narrative\_conflict (0.0--1.0): Does the content manufacture drama or conflict?}

\texttt{4. challenge\_format (0.0--1.0): Is this a challenge/dare/competition format?}

\texttt{5. commercial\_content (0.0--1.0): Is there visible product placement, brand logos, or sponsored content signals?}

\texttt{6. privacy\_violation (0.0--1.0): Does the content expose the child's private moments, body, or medical situations?}

\texttt{7. overall\_exploitative (0.0--1.0): Considering ALL dimensions, is this video likely exploitative?}

\texttt{Return ONLY valid JSON.}
\end{quote}

\end{document}